\documentclass[12pt]{article}
\usepackage{cite}
\usepackage{graphics}

\newcommand{\tb}{\ensuremath{\tan\beta}}
\newcommand{\stopp}{\ensuremath{\tilde t}}
\newcommand{\sbottom}{\ensuremath{\tilde b}} 
\newcommand{\cmin}{\ensuremath{\chi^-}}
\newcommand{\neut}{\ensuremath{\chi^0}}
\newcommand{\sg}{\ensuremath{\tilde{g}}}

\newcommand{\mw}{\ensuremath{M_W}}
\newcommand{\pl}{P_L}
\newcommand{\pr}{P_R}

\newcommand{\GeV}{\mbox{ GeV}}

\newcommand{\gsim}{\mbox{ \raisebox{-4pt}{${\stackrel{\textstyle >}{\sim}}$} }}
\newcommand{\lsim}{\mbox{ \raisebox{-4pt}{${\stackrel{\textstyle <}{\sim}}$} }}

\newcommand{\mg}{\ensuremath{m_{\tilde{g}}}}
\newcommand{\msba}{\ensuremath{m_{\tilde{b}_a}}}

\newcommand{\sbt}{{\tilde b}}
\newcommand{\higsi}{{\tilde h^-}}
 
\newcommand{\figsbdecaybr}{
  \begin{figure}[tb]
    \begin{center}
      \begin{tabular}{cc}
        \resizebox{7.0cm}{!}{\includegraphics{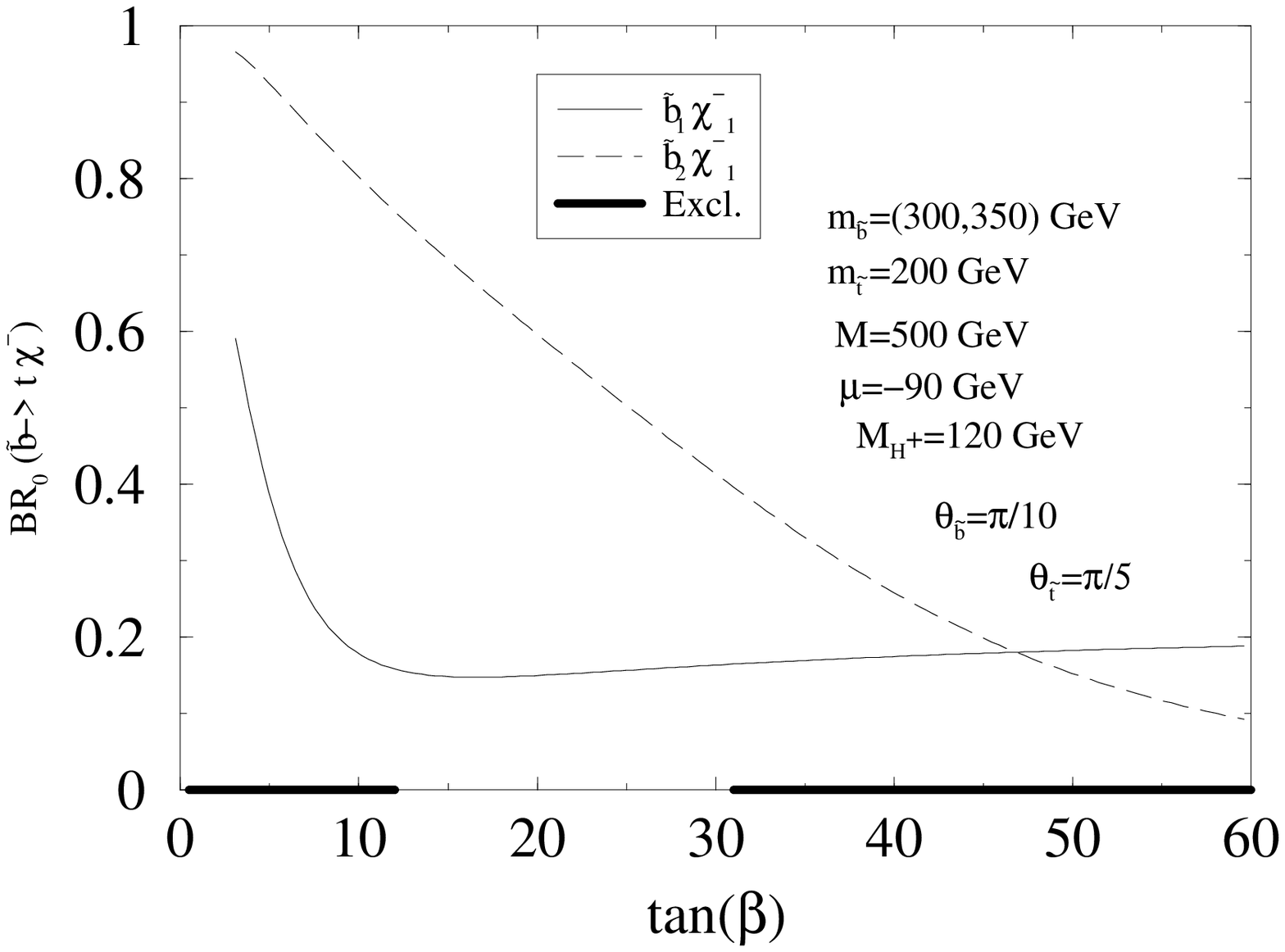}}&
        \resizebox{7.0cm}{!}{\includegraphics{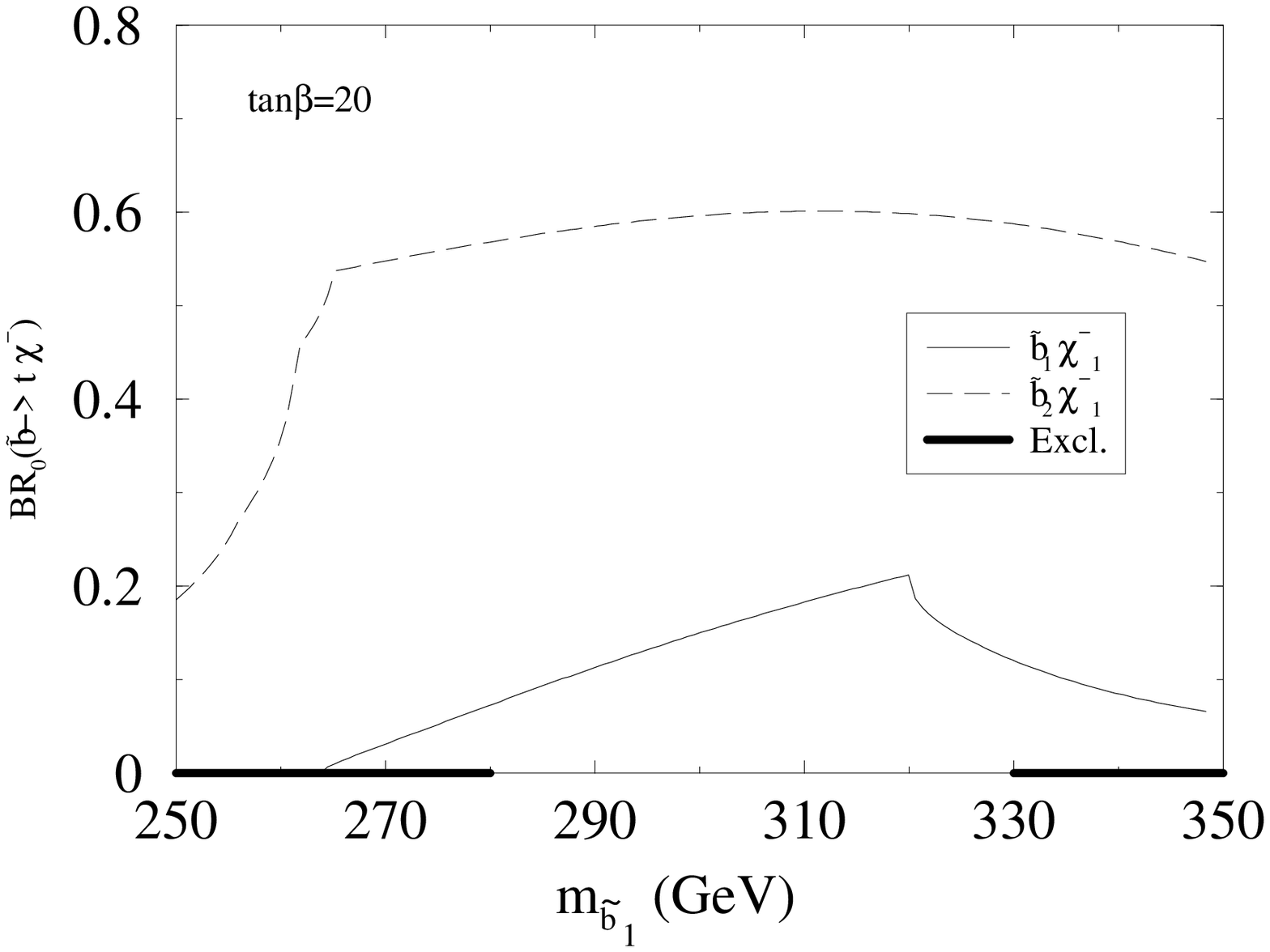}}
        \\
        (a) & (b) \\
        \multicolumn{2}{c}{\resizebox{7.0cm}{!}{\includegraphics{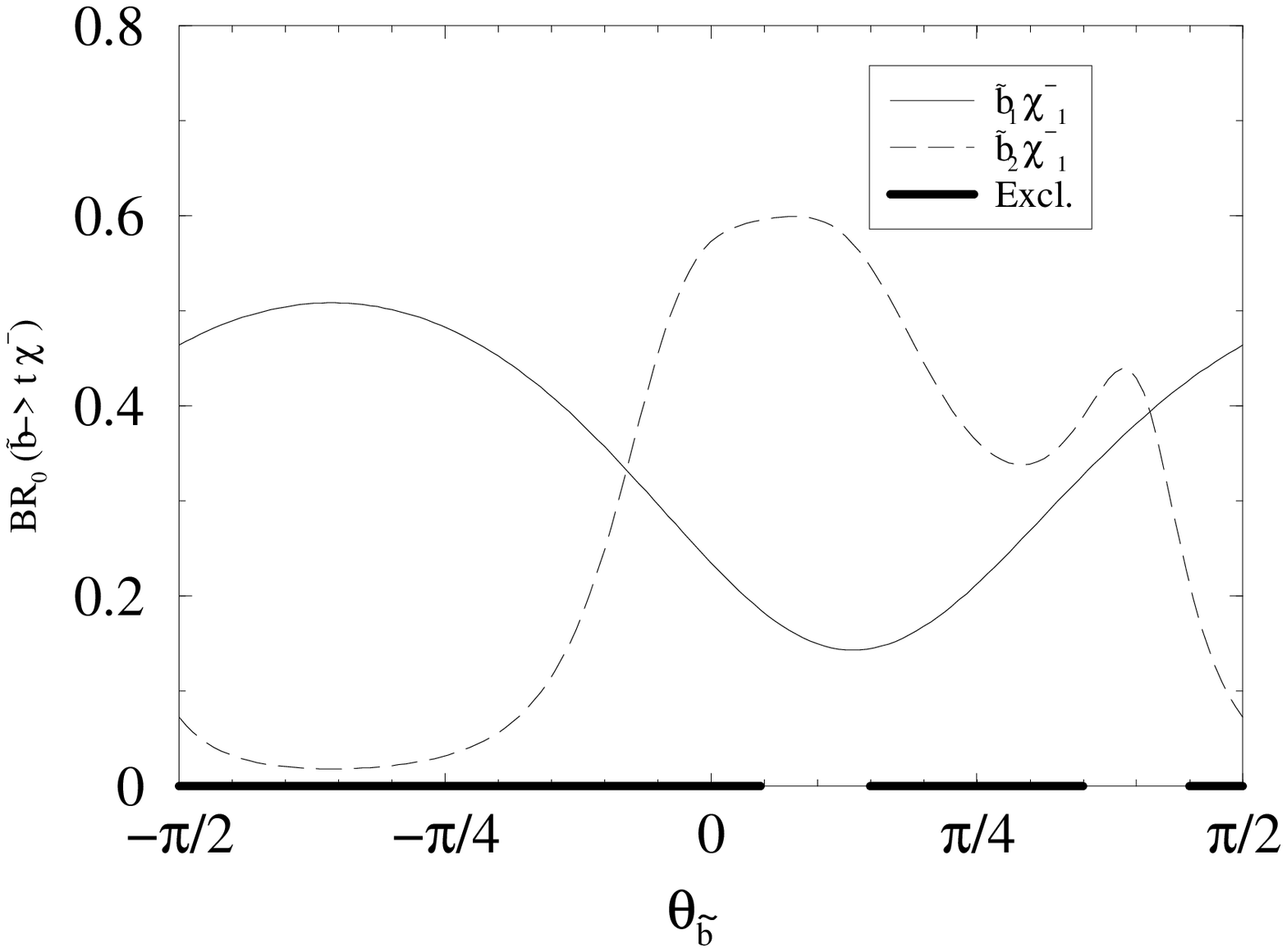}}}\\
        \multicolumn{2}{c}{(c)}
      \end{tabular}
    \end{center}
    \caption{{\bf (a)} The
      branching ratio of\,  
      $\tilde{b}_a\rightarrow\chi_1^-\,t$\, as a function of $\tan\beta$ for the
      various decays $a=1,2$ with $m_{\tilde{b}_1}<m_{\tilde{b}_2}$;
      {\bf (b)} As in (a), but as a function of $m_{\tilde{b}_1}$;
      {\bf (c)} As in (a), but as a function of $\theta_{\sbottom}$. 
      The marked parts of the abscissa in both figures are excluded
      by the condition (\protect\ref{eq:necessary}). The fixed
      parameters for (a) and (b) are given in the frame.\label{fig:sbdecaybr}
      }
  \end{figure}
  }
    
\newcommand{\figsbdecayQCDtbmu}{
  \begin{figure}[tb]
    \centerline{
      \begin{tabular}{c@{\hspace{1cm}}c}
        \resizebox{7cm}{!}{\includegraphics{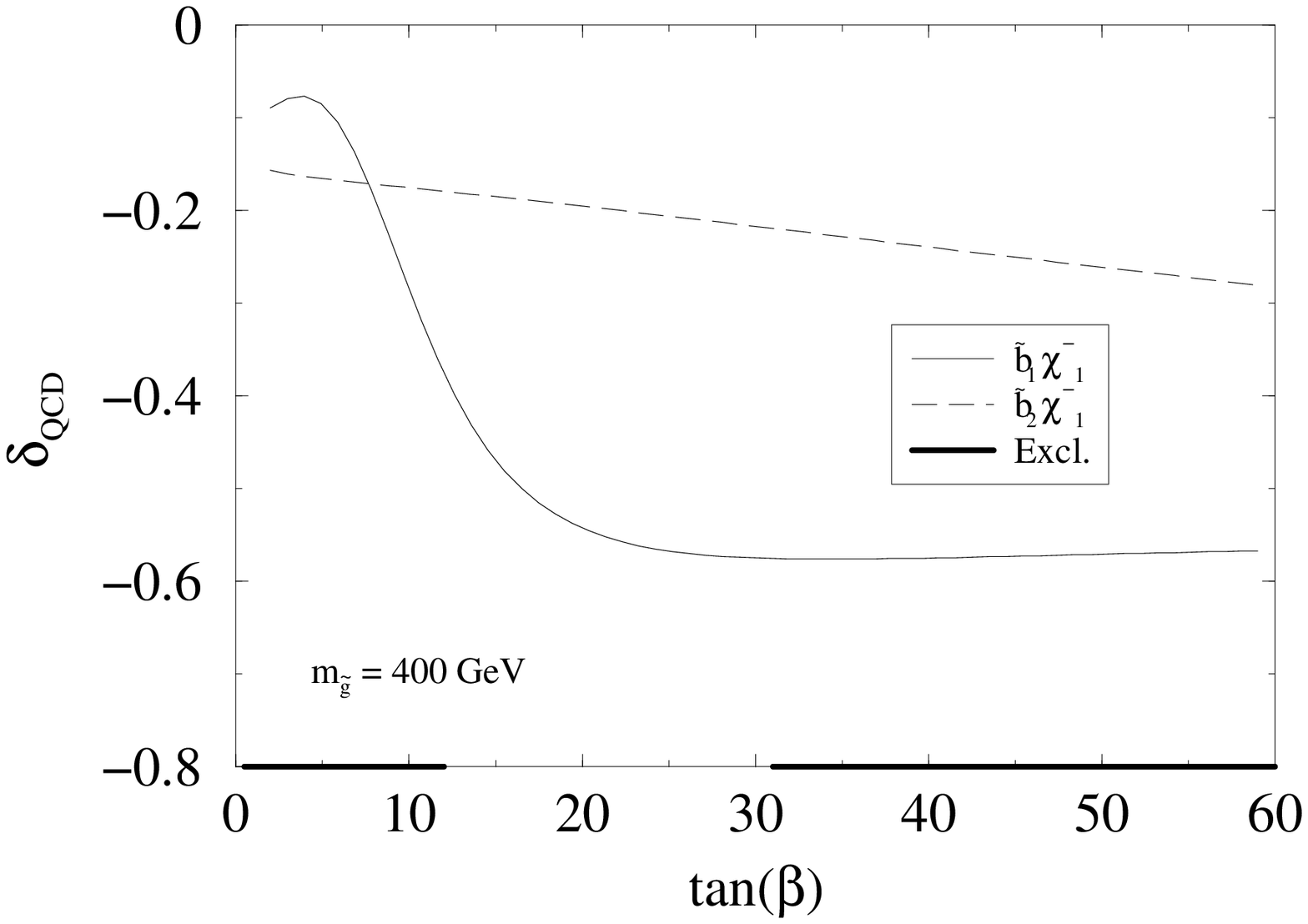}} &
        \resizebox{7cm}{!}{\includegraphics{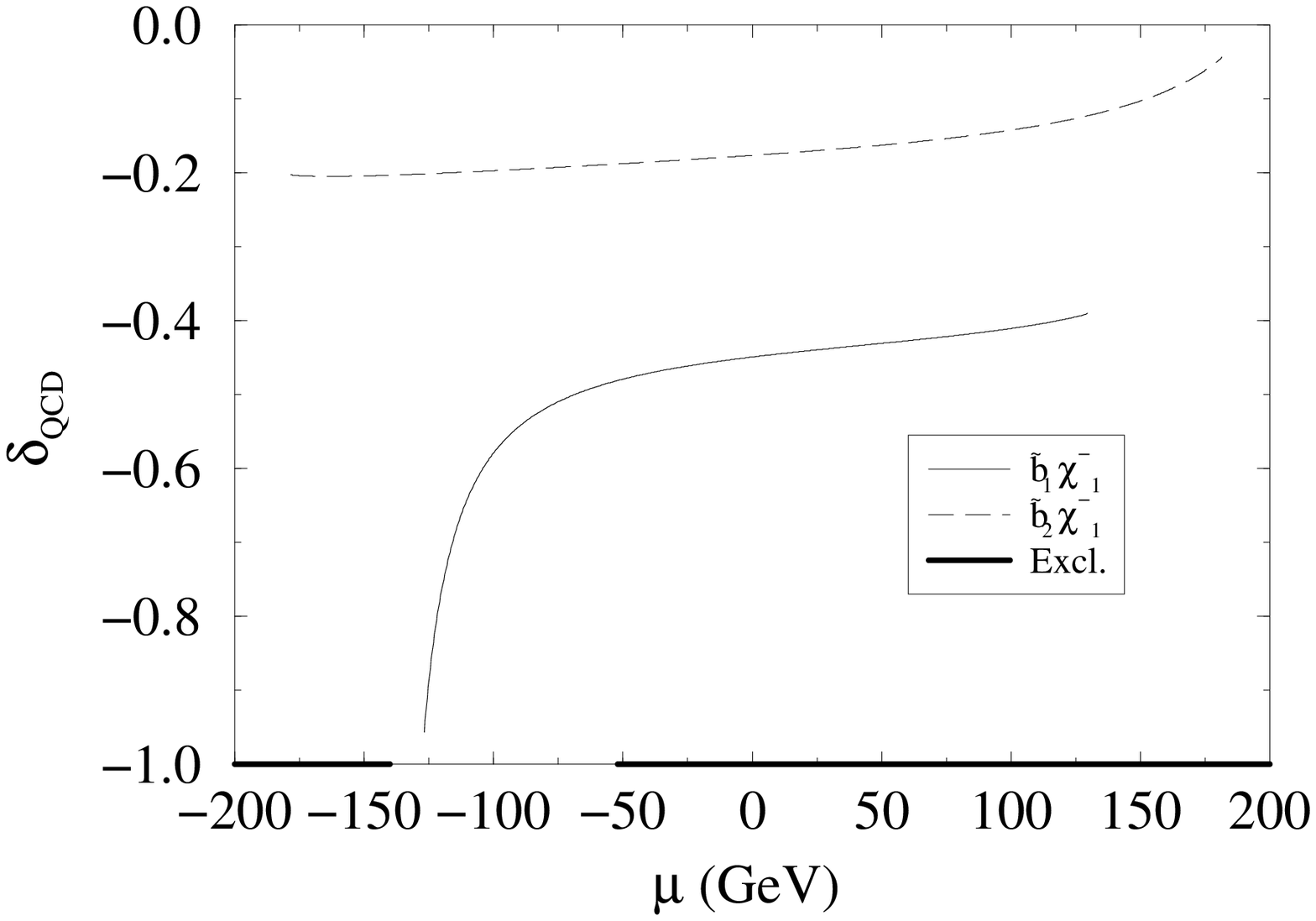}}\\
        (a)&(b)
      \end{tabular}
      }
    \caption{{The correction $\delta^{ai}_{QCD}$ to the decay width 
        $\sbottom_a\rightarrow t\,\cmin$ as a function of {\bf (a)} \tb\  and 
        {\bf  (b)} the higgsino mass parameter $\mu$. Set of inputs as in
        Fig.\,{\protect\ref{fig:sbdecaybr}}.}\label{fig:sbdecayQCDtbmu}}
  \end{figure}
  }

\newcommand{\figsbdecayQCDangle}{
  \begin{figure}[tb]
    \centerline{\resizebox{7cm}{!}{\includegraphics{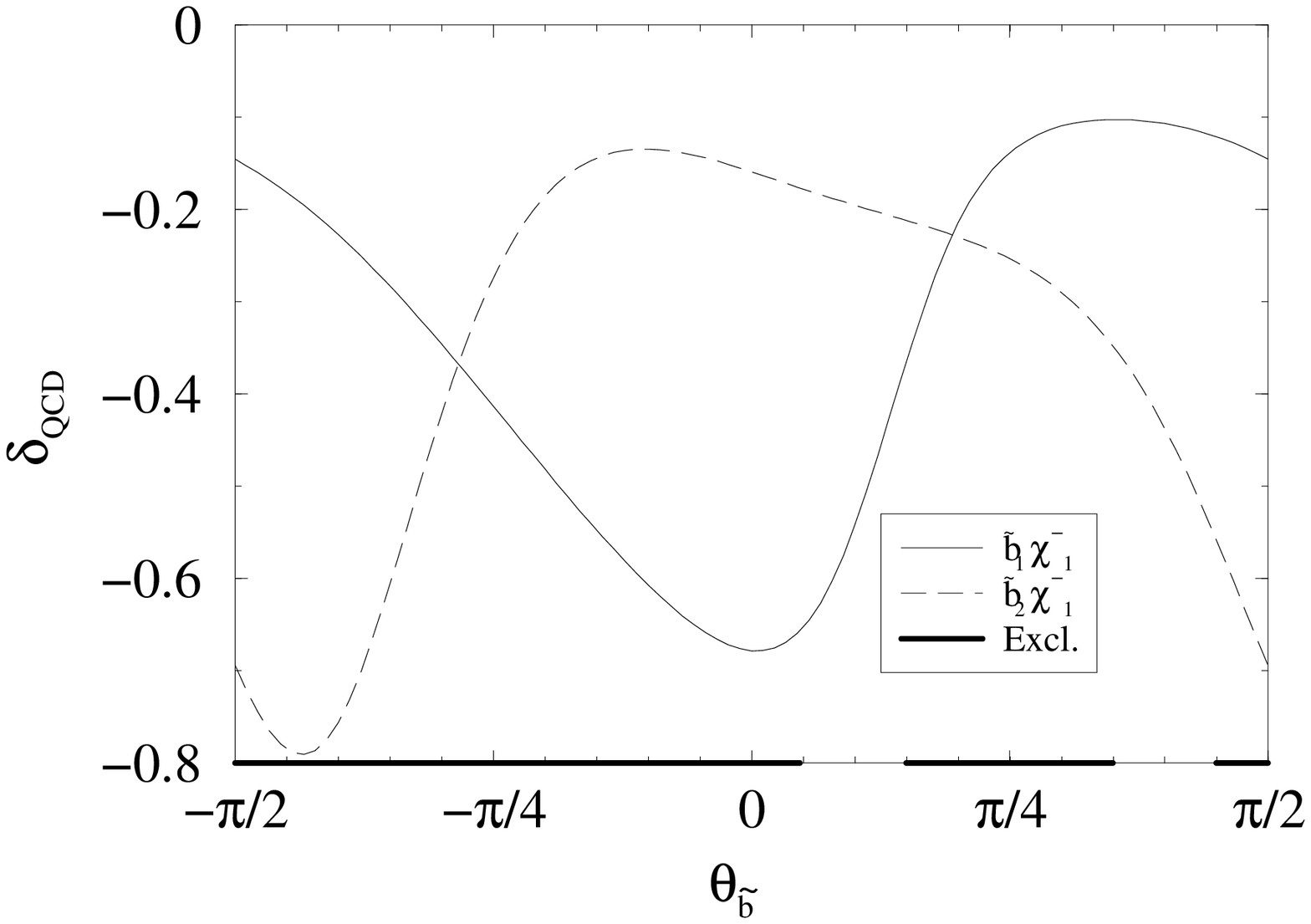}}}
    \caption{The correction $\delta^{ai}_{QCD}$ as a function of
      $\theta_{\sbottom}$. Inputs as in Fig.\,\protect{\protect\ref{fig:sbdecaybr}}.}
    \label{fig:sbdecayQCDangle}\end{figure}}

\newcommand{\figsbdecayQCDmgM}{
  \begin{figure}[tb]
    \centerline{
      \begin{tabular}{c@{\hspace{1cm}}c}
        \resizebox{7cm}{!}{\includegraphics{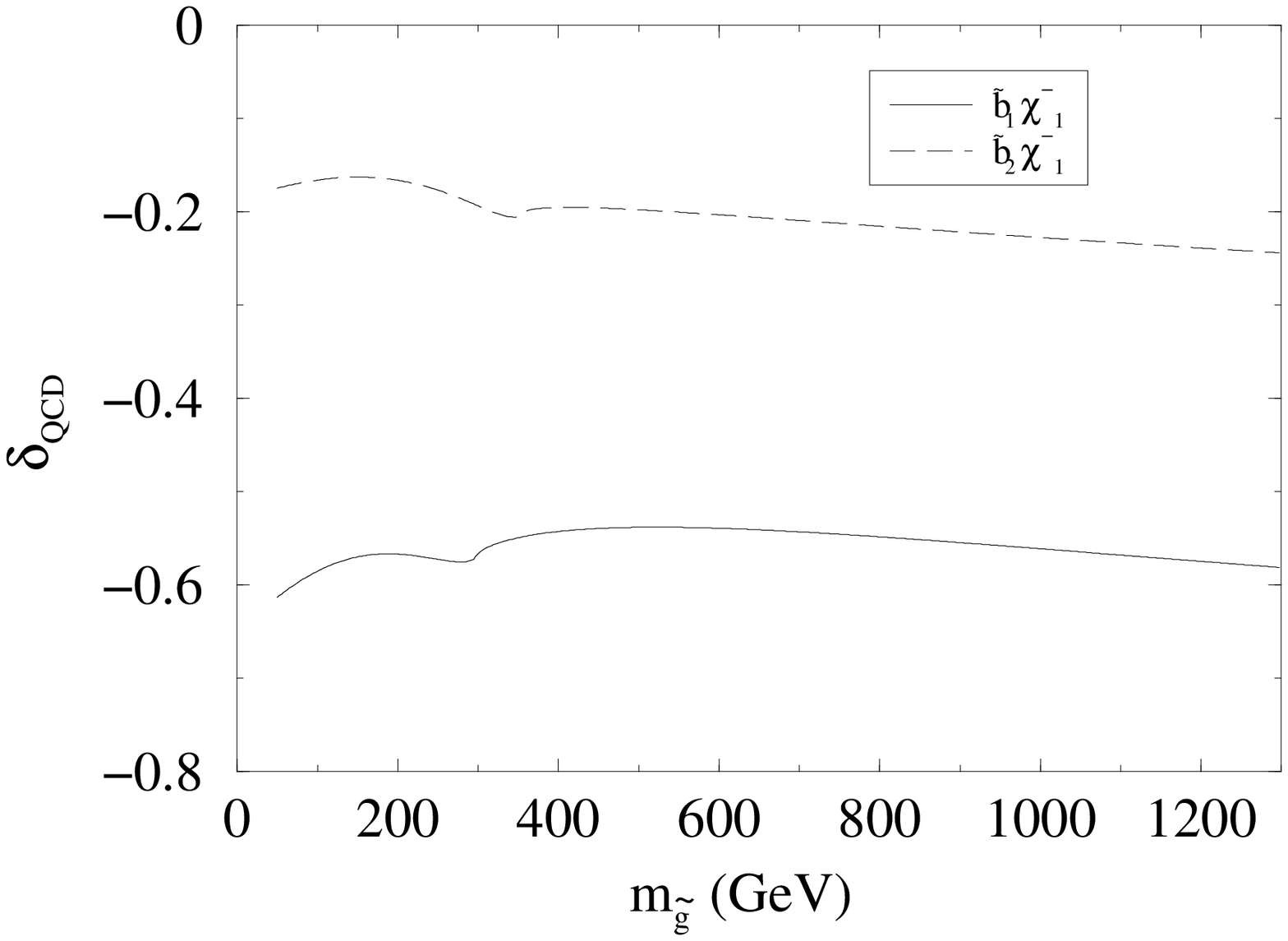}} &
        \resizebox{7cm}{!}{\includegraphics{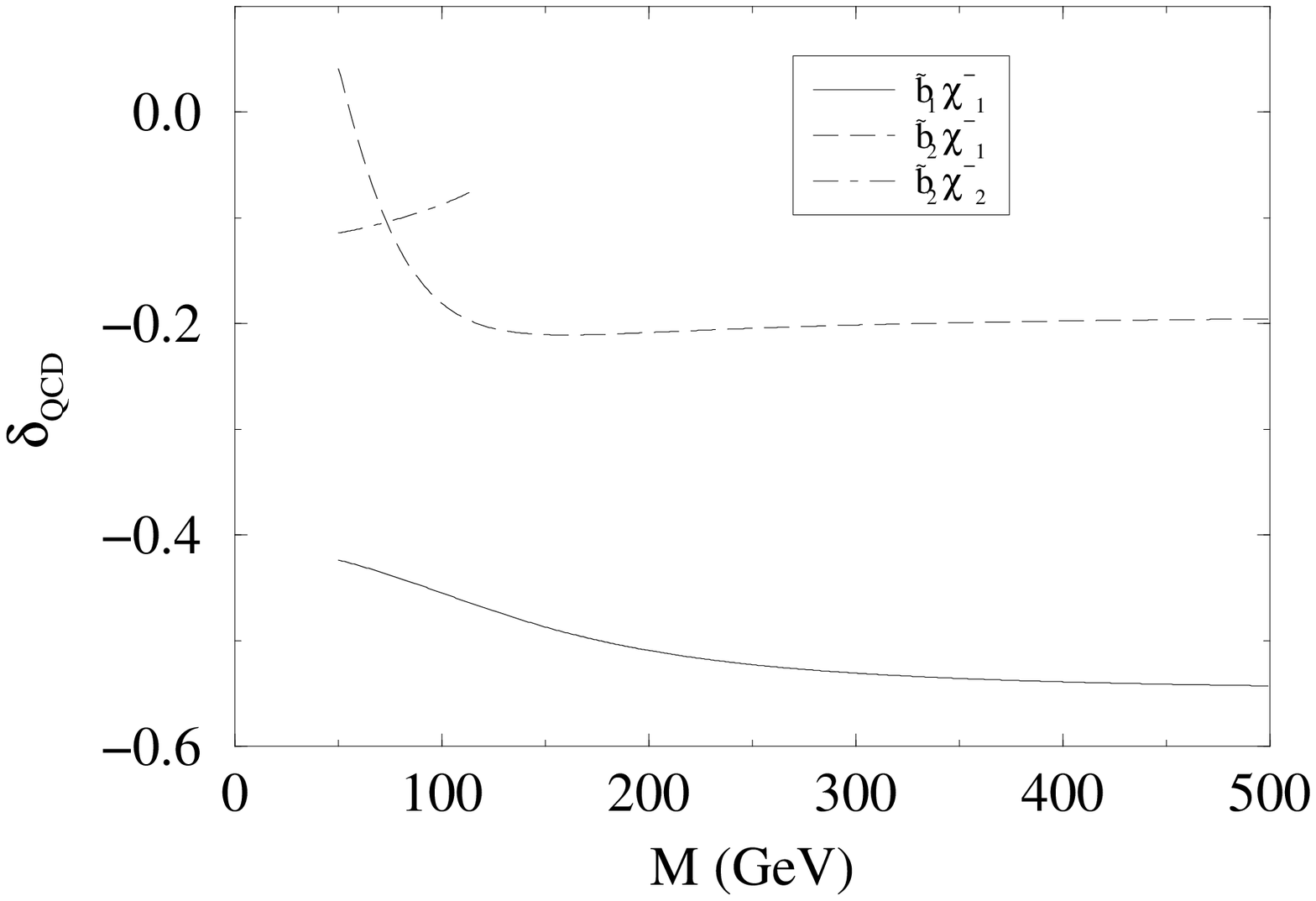}}\\
        (a)&(b)
      \end{tabular}
      }
    \caption{{The correction $\delta^{ai}_{QCD}$ as a function of {\bf (a)} the gluino mass
        \mg\  and 
        {\bf  (b)} the gaugino mass parameter $M$. Set of inputs as in
        Fig.\,\protect{\protect\ref{fig:sbdecaybr}}.}\label{fig:sbdecayQCDmgM}}
  \end{figure}}

\newcommand{\figsbdecaytres}{
  \begin{figure}[tb]
    \begin{center}
      \begin{tabular}{cc}
        \resizebox{7cm}{!}{\includegraphics{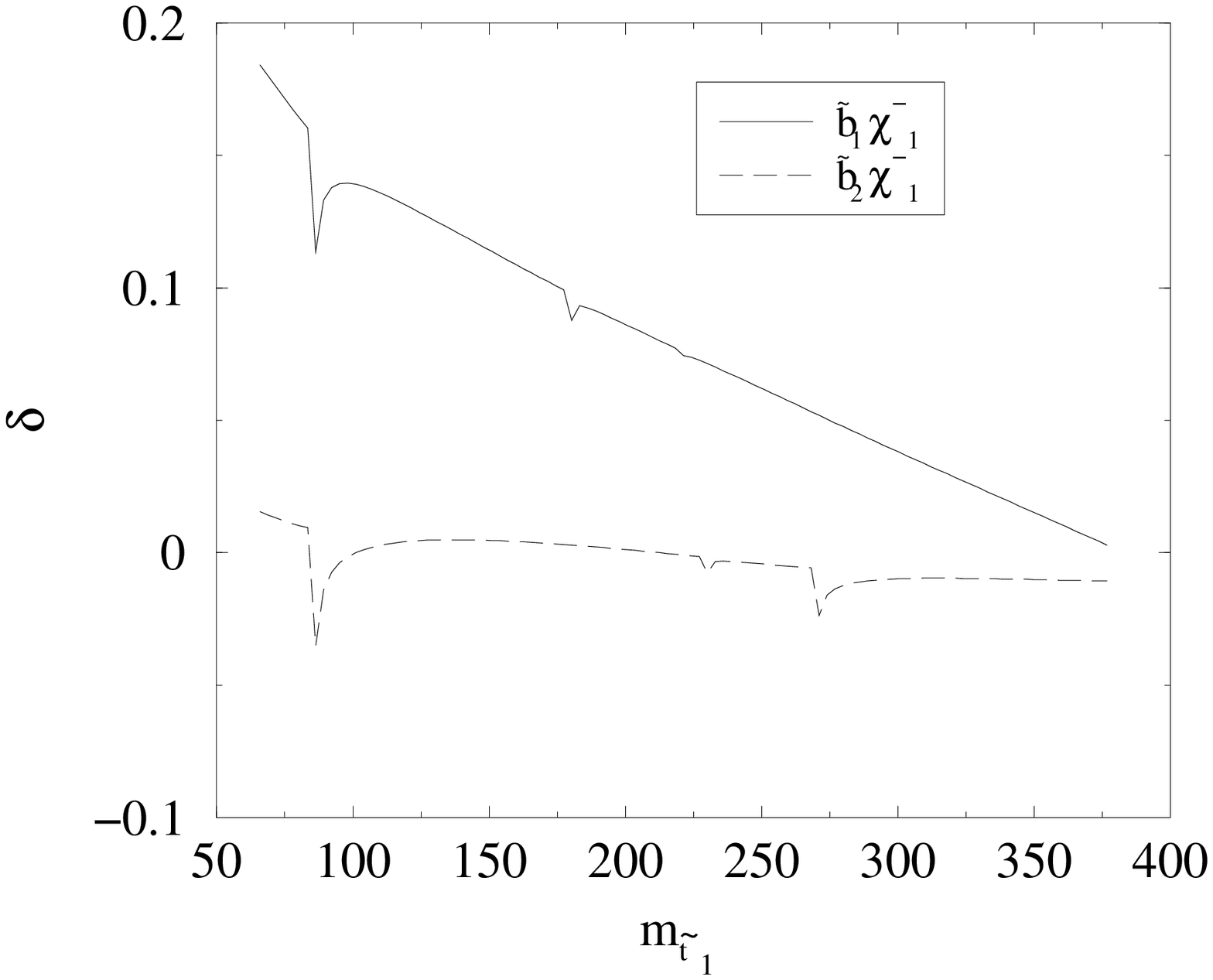}} &
        \resizebox{7cm}{!}{\includegraphics{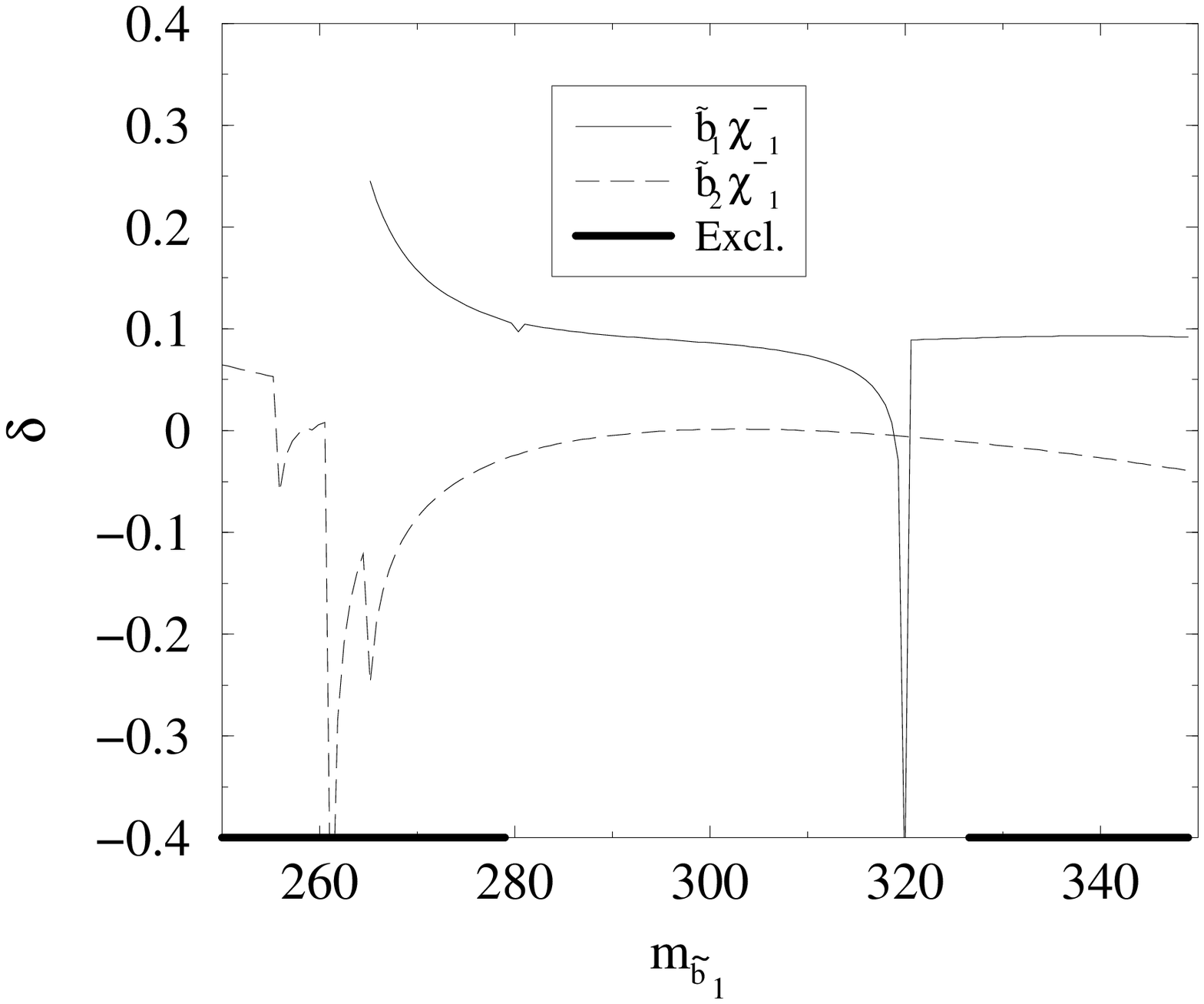}} \\
        (a) & (b) 
      \end{tabular}
      \caption{{\bf (a)} The SUSY-EW
        corrections~(\protect\ref{eq:sbtcharcorrection}) as a function of
        $m_{\tilde{t}_1}$; 
        {\bf (b)} As in (a), but as a function of $m_{\tilde{b}_1}$. 
        Rest of inputs as in Fig.\,\protect\ref{fig:sbdecaybr}.\label{fig:sbdecaytres}}
    \end{center}
  \end{figure}
  }

\newcommand{\figsbdecayquatre}{
  \begin{figure}[tb]
    \begin{center}
      \begin{tabular}{cc}
        \resizebox{7cm}{!}{\includegraphics{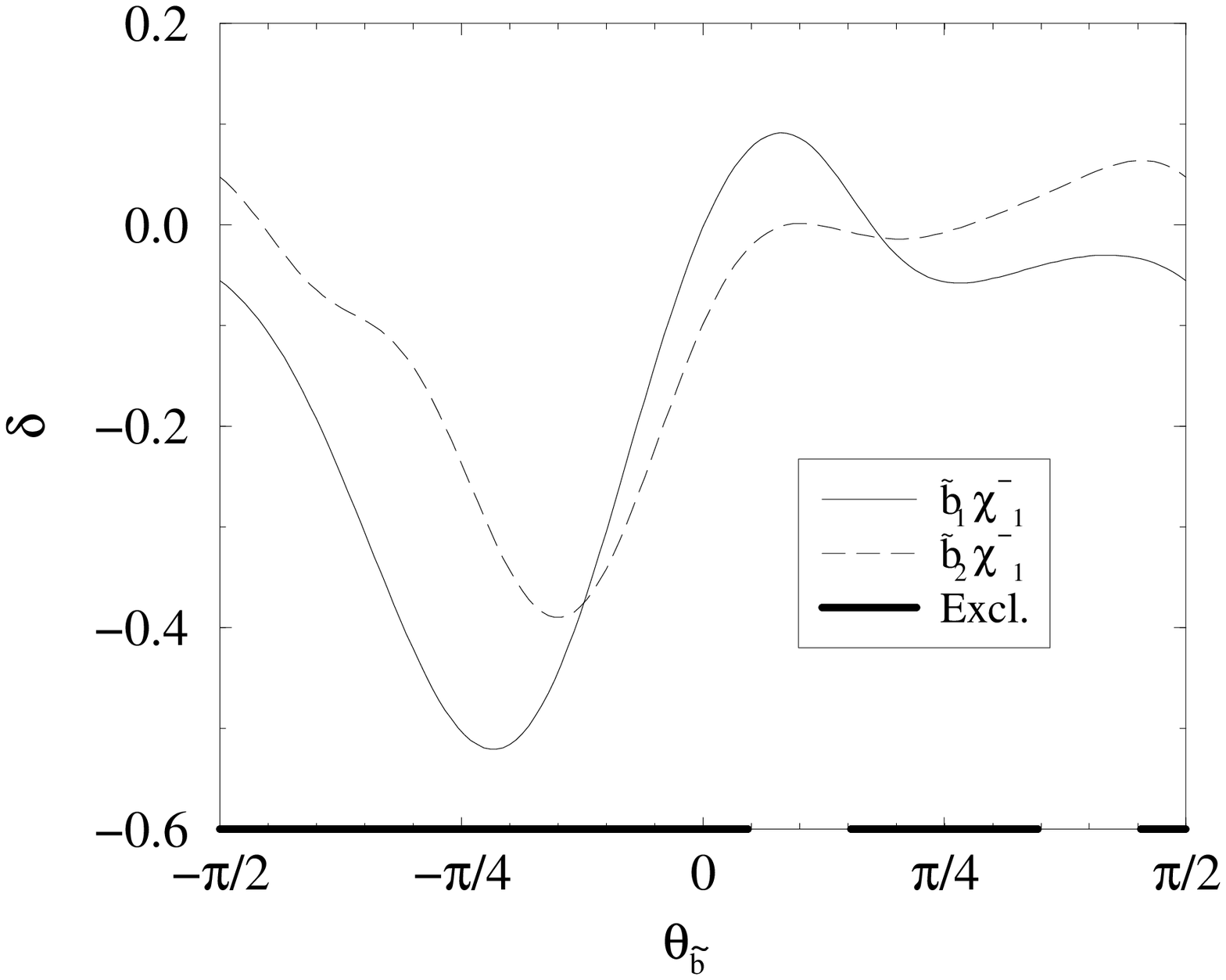}} &
        \resizebox{7cm}{!}{\includegraphics{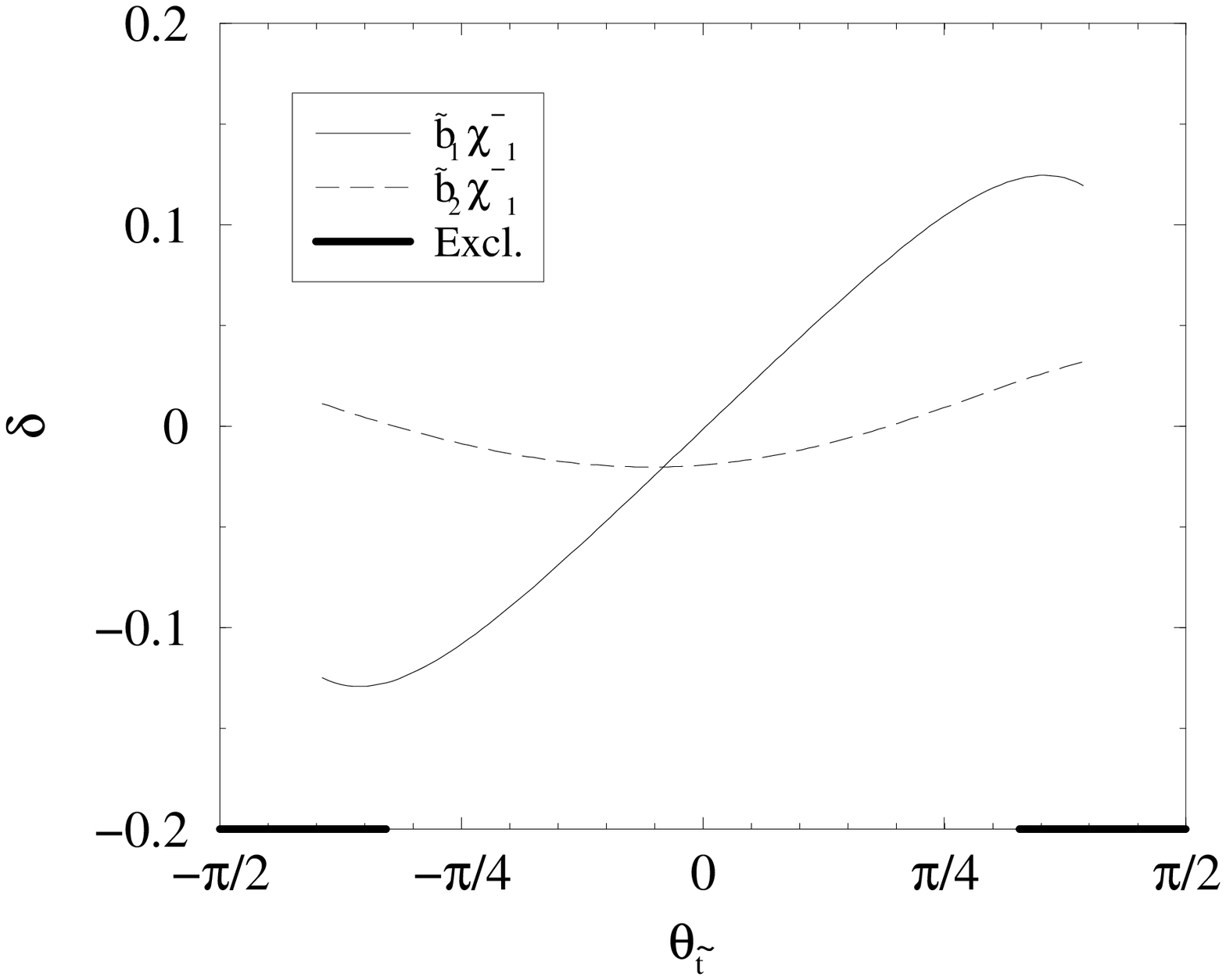}} \\
        (a) & (b)
      \end{tabular}
      \caption{{\bf (a)} Evolution of the SUSY-EW corrections as a
        function of the sbottom mixing angle, ${\theta}_{\tilde{b}}$, within 
        its allowed range;
        {\bf (b)} As in (a), but  as a function of the
        stop mixing angle, ${\theta}_{\tilde{t}}$. 
        Remaining inputs are as in Fig.\,\protect\ref{fig:sbdecaybr}.\label{fig:sbdecayquatre}}
    \end{center}
  \end{figure}
  }

\newcommand{\figsbdecaycinc}{
  \begin{figure}[tb]
    \begin{center}
      \begin{tabular}{cc}
        \resizebox{7cm}{!}{\includegraphics{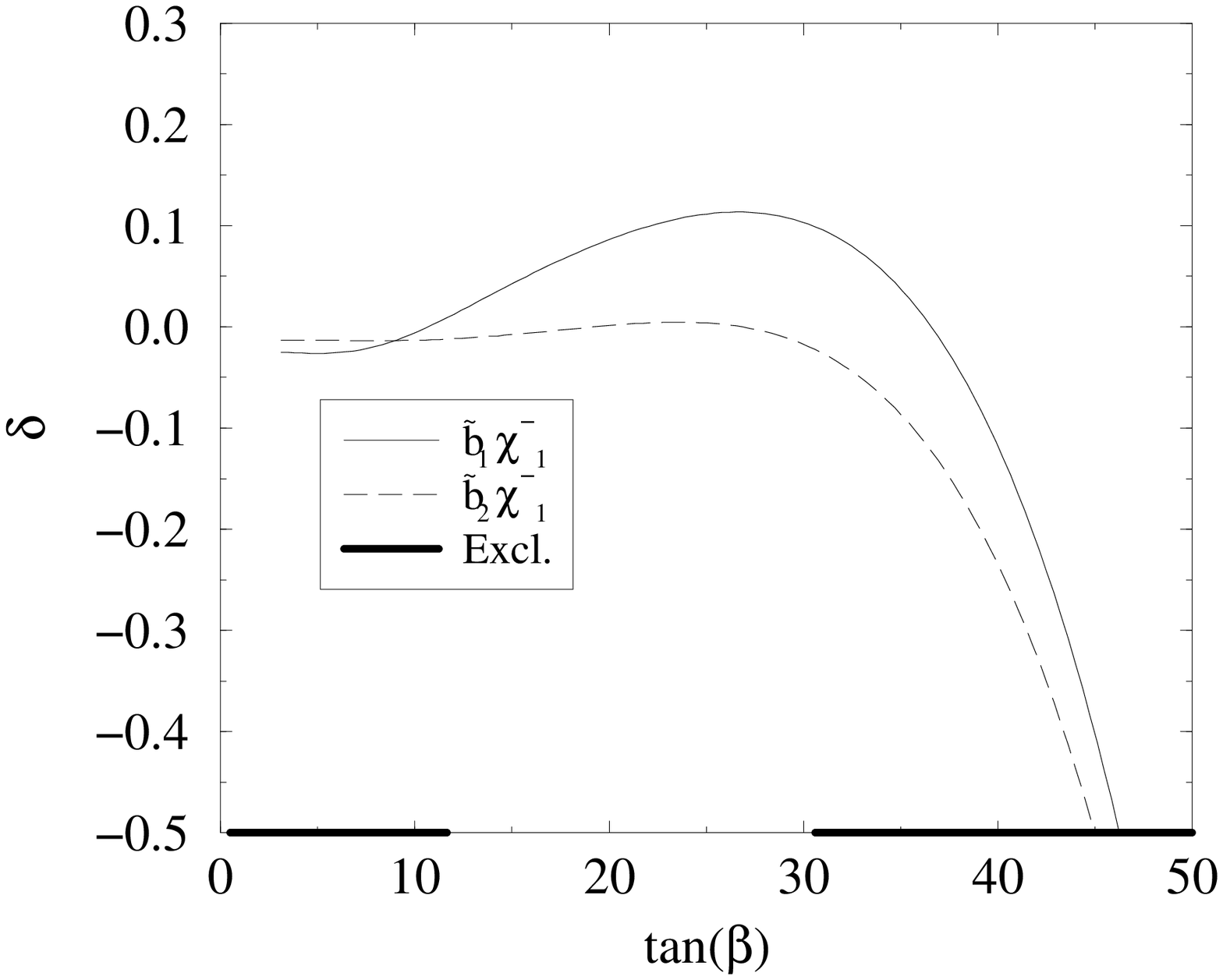}} &
        \resizebox{7cm}{!}{\includegraphics{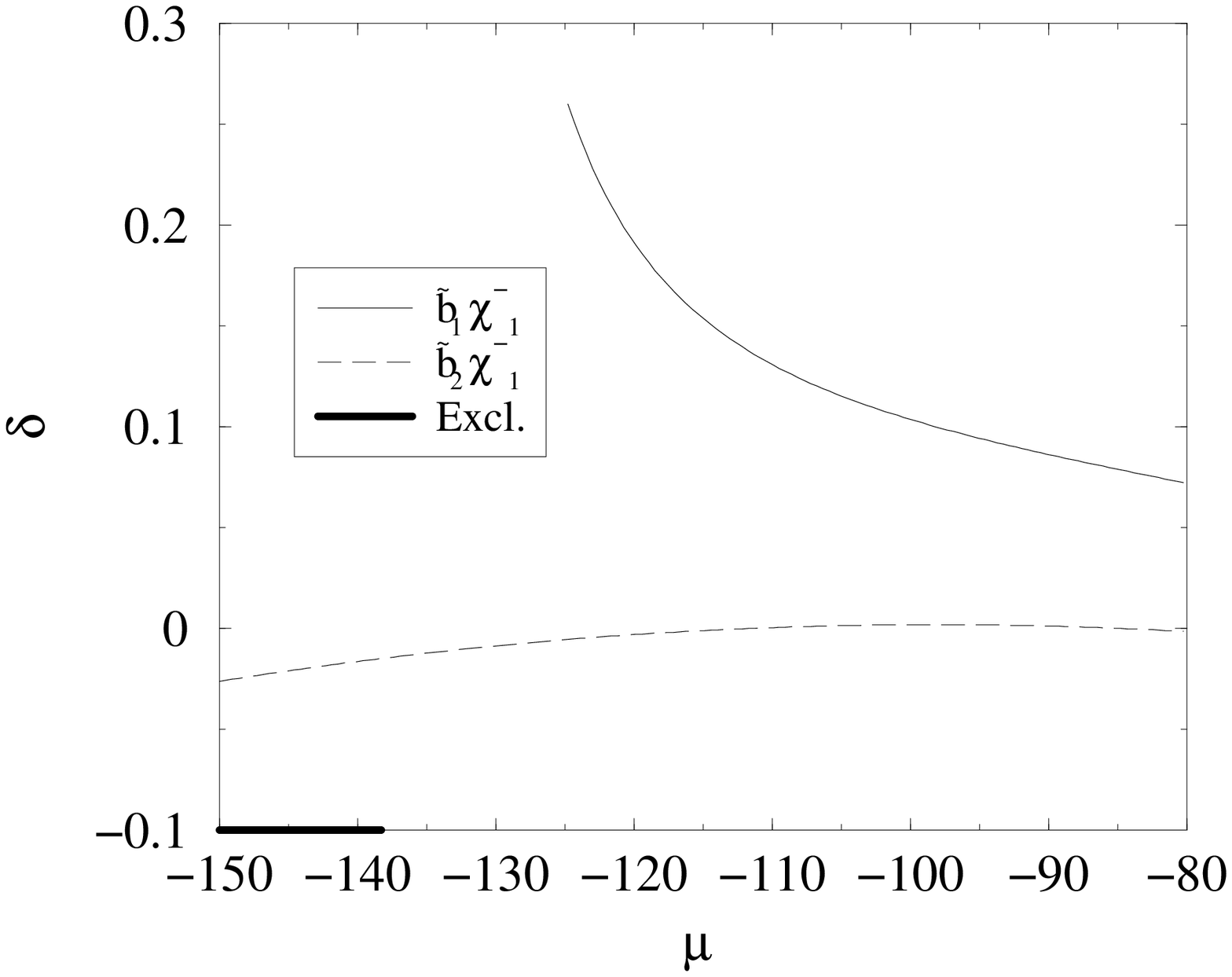}} \\
        (a) & (b)
      \end{tabular}
      \caption{{\bf (a)} The SUSY-EW correction as a function of 
        $\tan\beta$; {\bf (b)} As in (a), but as a function of $\mu$.
        Rest of inputs and notation as in
        Fig.\,\protect\ref{fig:sbdecaybr}.\label{fig:sbdecaycinc}}
    \end{center}
  \end{figure}
  }

\begin{document}

\thispagestyle{empty}
\begin{flushright}
{\parbox{3.5cm}{
UAB-FT-482\\ 
KA-TP-2-2000\\
LC-TH-2000-013\\
hep-ph/0001254\\
January 2000\\
}}
\end{flushright}
\vspace{3cm}
\begin{center}
{\Large \bf
Radiative Corrections to scalar quark decays in the MSSM\footnote{Talk presented
  by J. Guasch at the IIIth workshop in the 2nd ECFA/DESY Study
  on Physics and Detectors for a Linear Electron-Positron Collider,  Frascati
  (Italy) 7-10th November, 1998.}}
\vspace{1cm}
\vskip 8mm
{\large Jaume GUASCH, Wolfgang HOLLIK}

\medskip
{\sl Institut f\"{u}r Theoretische Physik, Universit\"{a}t Karlsruhe,}

{\sl  D-76128 Karlsruhe, Germany}
\bigskip

{\large Joan SOL{\`A}}\\
\medskip

{\sl Grup de F{\'\i}sica Te{\`o}rica and 
Institut de F{\'\i}sica d'Altes Energies}
 
{\sl Universitat Aut{\`o}noma de Barcelona
08193 Bellaterra (Barcelona), Catalonia, Spain}

\end{center}
\vspace{0.3cm}
\hyphenation{super-symme-tric co-lli-ders}
\hyphenation{com-pe-ti-ti-ve e-le-men-ta-ry}
\hyphenation{coun-ter-terms}
\begin{center}
{\bf ABSTRACT}
\end{center}
\begin{quotation}
\noindent
We review the one-loop corrections to the partial decay width of sbottom-quark
into a top quark and a chargino for the parameter space relevant to the TESLA
$e^+e^-$ linear collider. We present the results available in the literature for
the QCD and the Yukawa coupling corrections in a unified framework. In this way
a direct comparison of the size of the various corrections is possible.
\end{quotation}
  
\newpage

\section{Introduction}
Scalar quarks (the SUSY partners of quarks) could be produced at 
appreciable rates
at the TESLA collider. A large amount of work is already avaliable concerning the phenomenology
and the radiative corrections of the scalar fermion sector relevant for
the linear collider~(see
e.g.~\cite{Treeprod,Other,QCDprod,QCDdec,DHJ,KEBM,Yukprod,GSH} 
and references therein). 
With the experimental precision  expected at TESLA it will be become
necessary to include also the quantum corrections in the theoretical
investigations.
The largest radiative corrections for the squark sector of the MSSM
are associated with the
strong interaction.
Such QCD corrections 
have been investigated for the production
cross-section~\cite{QCDprod} and 
for several other squark observables~\cite{QCDdec,DHJ,KEBM}.
However, the electroweak (EW) corrections can also be sizeable and 
are in general not negligible.
This applies in particular to the
squarks of the third generation, owing to the large Yukawa couplings of the
fermions
\begin{equation}
\lambda_t\equiv {h_t\over g}={m_t\over \sqrt{2}\,M_W\,\sin{\beta}}\;\;\;\;\;,
\;\;\;\;\; \lambda_b\equiv {h_b\over g}={m_b\over \sqrt{2}
\,M_W\,\cos{\beta}}\,,
\label{eq:Yukawas} 
\end{equation}
where $\tb$ is the ratio between the vacuum expectation values of
the two Higgs bosons doublets, $\tb=v_2/v_1$. The Yukawa
couplings~(\ref{eq:Yukawas}) determine the strength of the interactions between
quarks and 
Higgs particles, squarks and higgsinos (the SUSY partners of the quarks and
Higgs particles respectively), and part of the interactions between the squarks
and Higgs bosons. 
However, in the case of squarks there are additional
interactions originating from the breaking of SUSY, the so-called soft-SUSY-breaking
trilinear terms. Although these terms are bounded by the condition that the
vacua do not break charge and color, they can be large enough to provide large
quantum corrections.
Recently, the Yukawa corrections to the production
cross-section~\cite{Yukprod}  have received attention.

We have computed the leading electroweak corrections to the partial decay width
of a bottom squark into a top quark and a chargino 
$\Gamma(\sbottom\to t\cmin)$~\cite{GSH}. The reasons to choose this concrete channel are
at least twofold. First of all it is a process of special interest. The third
generation squarks can be the lightest sfermions of the model, owing to the large
Yukawa couplings that make their mass decrease  when one assumes 
a common sfermion mass
scale at a unification scale, according to the Renormalization Group
evolution. Secondly, the third generation squarks are most likely to develop
large EW 
corrections, owing to their large couplings to the Higgs sector. Of course, the
neutral channels ($\stopp\to t\neut$ and $\sbottom \to b\neut$) are equally
interesting. 

A key point in the computation of observables with $R$-odd external particles is
that it is no longer possible to separate between SUSY and non-SUSY
corrections. This  means that, when making the appropriate renormalization,
both the $R$-even and $R$-odd sectors of the theory have to be renormalized. Hence, for
the computation of the decay process mentioned above we should perform the
renormalization of all the neutralino-chargino sector, together with the gauge
and the Higgs sector, which is a rather voluminous task. We have therefore
chosen a scenario that allows a simplified treatment: If we assume that the
gaugino soft-SUSY-breaking masses are 
much larger than the higgsino mass parameter $|\mu|$ then we can treat the
$\cmin$ appearing in the process as a purely higgsino particle; in this way
we can avoid to deal with all the plethora of gauge and gaugino particles in the
electroweak sector. Of
course this is only a first approximation, but it is already sufficient to demonstrate the
importance of the corrections, before performing  the full
computation. This specification to the leading Yukawa terms is 
meaningful only in the higgsino approximation, where the soft-SUSY-breaking gaugino
masses have to fulfill the relation
$$
M',M \gg \{|\mu|,\mw\}\,\,,
$$
with the lightest chargino as pure higgsino, that is
\begin{equation}
\cmin_1 = \higsi\,\,,\,\,m_{\cmin_1}\simeq|\mu|\,\,,\,\,\cmin_2 =
\tilde{w}^-\,\,,\,\,m_{\cmin_2}\simeq M\,\,.
\label{eq:defhigsi}
\end{equation}

\section{Tree-level relations}
The tree-level Lagrangian for the top-sbottom-chargino interactions reads
\begin{equation}
  {\cal L}_{t\sbt\chi}=-g\, \sbt^*_a 
\bar{\chi}^+_i (A_{+}^{ai}\pl+\epsilon_i\, 
A_{-}^{ai}\pr) t +{\rm h.c.}\,,
\label{eq:ltl}
\end{equation}
where $\epsilon_i$ is the sign of the {\it i}th chargino eigenvalue $M_i\,
(i=1,2$ with $|M_1|<|M_2|$) in the real matrix representation, and the coupling
matrices are denoted by\footnote{See Refs.\cite{CGGJS,Guasch1,JaumeTes} for
full notation niceties.}
\begin{equation}
  A_{+}^{ai}=R_{1a} V_{i1}-\lambda_b R_{2a} V_{i2}\,,\ \ \ \
  A_{-}^{ai}=-R_{1a} \lambda_t U_{i2}\,.
\label{eq:AS}
\end{equation}
The explicit appearance of the Yukawa couplings~(\ref{eq:Yukawas})
in the Lagrangian above requires both the
introduction of top and bottom quark mass counterterms
(in the on-shell scheme) and also a suitable prescription for 
the renormalization of \tb. 
We denote by $m_{\tilde{b}_a}\,(a=1,2)$, with
$m_{\tilde{b}_1}<m_{\tilde{b}_2}$, the two sbottom mass eigenvalues.
The sbottom mixing angle $\theta_{\tilde{b}}$
is defined by the transformation relating the weak-interaction
($\sbt^\prime_a=\sbt_L, \sbt_R$) and the mass eigenstate
 ($\sbt_a=\sbt_1, \sbt_2$) squark bases:
\begin{equation}
\label{eq:defsq}
  \sbt^\prime_a=R_{ab}\, \sbt_b\,\,; \,\,\,\,
  R=\left(\begin{array}{cc}
      \cos\theta_{\tilde{b}}&-\sin\theta_{\tilde{b}} \\
      \sin\theta_{\tilde{b}}&\cos\theta_{\tilde{b}}
    \end{array}\right)\,;
\end{equation}
$R$ is the matrix appearing in eq.~(\ref{eq:AS}).
By this  basis transformation, the sbottom mass matrix,
\begin{equation}
{\cal M}_{\tilde{b}}^2 =\left(\begin{array}{cc}
M_{\tilde{b}_L}^2+m_b^2+\cos{2\beta}(-{1\over 2}+
{1\over 3}\,s_W^2)\,M_Z^2 
 &  m_b\, (A_b-\mu\tan\beta)\\
m_b\, (A_b-\mu\tan\beta) &
M_{\tilde{b}_R}^2+m_b^2-{1\over 3}\,\cos{2\beta}\,s_W^2\,M_Z^2\,,  
\end{array} \right)\,,
\label{eq:sbottommatrix}
\end{equation}
becomes diagonal: 
$R^{\dagger}\,{\cal M}_{\tilde{b}}^2\,R=
{\rm diag}\left\{m_{\tilde{b}_2}^2, m_{\tilde{b}_1}^2\right\}\,.
$

Our aim is to compute the radiative corrections in an on-shell
renormalization scheme; hence, the input parameters are  physical observables (i.e.\ the physical masses
$m_{\tilde{b}_2}, m_{\tilde{b}_1}$, \ldots) rather than formal parameters in the Lagrangian
(i.e.\ the soft-SUSY-breaking parameters $M_{\tilde{b}_L}^2, A_b$, \ldots\ in
eq.~(\ref{eq:sbottommatrix})). Specifically, we  use the following set of
independent parameters for the squark sector:
\begin{equation}
(m_{\tilde{b}_1}, m_{\tilde{b}_2}, \theta_{\tilde{b}},m_{\tilde{t}_1}, \theta_{\tilde{t}})\,.
\label{eq:inputb}\label{eq:inputt}
\end{equation}
The value of the other
stop mass $m_{\tilde{t}_2}$ is then determined by $SU(2)_L$ gauge invariance. 
For the numerical study, we
shall use a range of 
bottom-squark masses $300-350\GeV$, relevant for a $\sqrt{s}=800\GeV$ $e^+e^-$
linear collider.
The sbottom and stop trilinear soft-SUSY-breaking terms $A_b$ and $A_t$ are fixed at the
tree-level by the previous parameters as follows:
\begin{equation}
A_{b}=\mu\,\tan\beta+
{m_{\tilde{b}_2}^2-m_{\tilde{b}_1}^2\over 2\,m_b}\,\sin{2\,\theta_{\tilde{b}}}\,;
\ \ \ \
A_{t}=\mu\,\cot\beta+
{m_{\tilde{t}_2}^2-m_{\tilde{t}_1}^2\over 2\,m_t}\,\sin{2\,\theta_{\tilde{t}}}\,.
\label{eq:Abt}
\end{equation}
We impose the approximate (necessary) condition
\begin{equation}
A_q^2<3\,(m_{\tilde{t}}^2+m_{\tilde{b}}^2+M_H^2+\mu^2)\,,
\label{eq:necessary}
\end{equation}
where $m_{\tilde{q}}$ is of the order of the average squark masses
for $\tilde{q}=\tilde{t},\tilde{b}$, to avoid colour-breaking minima 
in the MSSM Higgs potential\,\cite{Frere}. Of course the relation~(\ref{eq:Abt})
receives one-loop corrections. However, since these parameters do not enter the
tree-level expressions, these effects translate into two-loop corrections to the
process under study. The bound~(\ref{eq:necessary}) translates into a stringent
constrain to the sbottom-quark mixing angle for moderate and large values of
$\tb \gsim 10$: with an approximate limit $|\mu|\gsim 80\GeV$  from the
negative output of the chargino search at LEP,  the
condition~(\ref{eq:necessary}) can only be satisfied by a cancellation of the
two terms in~(\ref{eq:Abt}) which is easily spoiled when $\theta_\sbt$ is
varied. The right hand side of eq.~(\ref{eq:necessary}) is not rigorous; so we
will present results also when this bound is not satisfied, but we will clearly
mark these regions. With the use of the bound~(\ref{eq:necessary}) also the
squark-squark-Higgs-boson couplings are restricted. This is a welcome feature,
since these couplings can 
in general be very large, eventually spoiling perturbativity.

It is clear that the radiative corrections to the process 
$\sbottom_a\rightarrow t\,\cmin_i$ will only be of practical  interest in the
region where it 
also has a large tree-level branching ratio. 
There are several channels ($\sbottom_a\to b\sg$, $\sbottom_a\rightarrow
b\,\neut_\alpha$, $\sbottom_2\rightarrow \sbottom_1 h^0$,~\ldots) that 
contribute to the sbottom-quark decay width. The gluino channel, if
kinematically avaliable, saturates the total width, so in order 
to have an appreciable branching ratio $\sbottom_a \rightarrow t\,\cmin_i$ we
start out assuming that the gluino is much heavier than the squarks
$\mg > \msba$, $a$=1, 2.
Neutralino masses, on the other hand, are related to chargino masses; thus, no
additional conditions can be imposed on this side. 

Let us define the branching ratio for the decay under investigation:
\begin{equation}
BR_0(\tilde{b}_a\rightarrow t\,\cmin_1)=
{\Gamma_0(\tilde{b}_a\rightarrow
  \cmin_1\,t)\over\Gamma_0^T(\sbottom_a)}\,\,,
\label{eq:sbdecayBR}
\end{equation}
where $\Gamma_0^T(\sbottom_a)$ is the total $\sbottom_a$ decay width.
This branching ratio is maximized
in a scenario  
where the lightest chargino is higgsino-dominated and \tb\  is of
low--moderate value. For large  $\tb\gsim 40$,  $\Gamma_0^T$ is dominated
by the neutral higgsino contribution.

\figsbdecaybr

Figure~\ref{fig:sbdecaybr} displays  the
value of the branching ratio~(\ref{eq:sbdecayBR}) as a function of \tb,
$m_{\sbottom_1}$ and $\theta_{\sbottom}$, for given values of the other
parameters. It  can be seen that low \tb\  values enhance the branching ratio. From now on we will
concentrate in the region of $\tb \simeq 20$; with this typical value the
branching ratio still is appreciably high, whereas the electroweak
corrections can be enhanced by means of the bottom Yukawa
coupling~(\ref{eq:Yukawas}). In Fig.\,\ref{fig:sbdecaybr}(b) we can see the
thresholds for opening
the Higgs channels, namely $\sbottom_2\rightarrow\sbottom_1\,h^0$ (at
the left end of the figure) and $\sbottom_1\rightarrow\stopp_1\,H^-$ (at
its right end). 
When these channels are open, they tend to decrease the branching
ratio~(\ref{eq:sbdecayBR}) to undetectable small values. 
The large decay
width into Higgs bosons results from large values for 
the $A$ parameters~(\ref{eq:Abt})
in these kinematical region. Of course one could fix the input
parameters~(\ref{eq:inputb}) in such a way that $A_{\{t,b\}}$ 
are small in one of
these regions (say at $m_{\sbottom_1}$ light), but at the price of making them
large at its central value and even larger at the other end. 
This effect is also
seen in Fig.\,\ref{fig:sbdecaybr}(c), as the $A$ parameters are related to the
angle trough~(\ref{eq:Abt}). 
Note that the allowed range of
$\theta_{\sbottom}$ is rather narrow, so that the physical bottom squark
mass eigenstates basically coincide with
the left- and right-handed chiral electroweak eigenstates.

\section{One-loop corrections}
The QCD one-loop corrections were originally computed in
Refs.\cite{DHJ,KEBM}. Our QCD results presented here were computed
independently and are in full agreement with those 
of~\cite{DHJ,KEBM}. We include them in our discussion, for comparison with the
residual ones, within the same scenario in which we computed the Yukawa
part\footnote{Our computation of the QCD effects can be found
in~\cite{JaumeTes}.}. 
The Yukawa corrections were first presented in~\cite{GSH}\footnote{The results presented here differ slightly from those of Ref.\cite{GSH} due to 
a recently discovered computer bug.}. The
full analytical results of the corrections can be found
in~\cite{DHJ,KEBM,GSH}. The QCD corrections contain all the gluon
and gluino exchange diagrams, together with the soft and hard gluon
bremsstrahlung, and the Yukawa corrections contain the diagrams in which Higgs
bosons and higgsinos are exchanged. We use the on-shell renormalization scheme
with the input parameters described in~(\ref{eq:inputb}). 
The renormalization of the
$\tb$ parameter (necessary for the weak corrections) is fixed in such a way 
that
the decay width $\Gamma(H^+\to\tau^+\nu_\tau)$ does not receive quantum
corrections~\cite{CGGJS}. The $\mu$ parameter is renormalized in analogy
to fermion mass renormalization, 
since in our approximation it is the mass of the chargino
involved in the decay. The bottom-squark mixing angle has to be
renormalized as well.
At variance with the other parameters appearing in our process, it is
still not clear how this angle could be measured.\footnote{For the top-squark mixing angle, a recent study has shown
that a good precision can be obtained  by
measuring the production cross-section $\sigma(e^+e^-\to \stopp_a\stopp_b)$, 
using polarized electrons, with the help of the polarization
asymmetry~\cite{Treeprod}.} 
Hence we treat $\theta_{\sbottom}$ as a formal parameter
and impose as a renormalization condition
that it is not shifted by loop corrections from
the mixed $\sbottom_1\sbottom_2$~self-energy~\cite{GSH}\footnote{Several
different renormalization conditions for the squark mixing angle have been
discussed in the literature, see e.g.~\cite{QCDprod,QCDdec,KEBM,DHJ,Yukprod,Plehn} and references therein.}.

The quantity under study will be the relative one-loop
correction defined as:
\begin{equation}
  \label{eq:sbtcharcorrection}
  \delta^{ai}=\frac{\Gamma(\sbottom_a\to t \cmin_i)-\Gamma_0(\sbottom_a\to t \cmin_i)}{\Gamma_0(\sbottom_a\to t \cmin_i)}\,\,.
\end{equation}
\figsbdecayQCDtbmu
\figsbdecayQCDangle
\figsbdecayQCDmgM
We start with the QCD corrections shown in Figs.~\ref{fig:sbdecayQCDtbmu}-\ref{fig:sbdecayQCDmgM}.
For
the numerical evaluation we use $\alpha_s(\msba)$, using the one-loop MSSM
$\beta$-function, but, for the $\msba$ we use, it is basically the 4-flavour SM
$\beta$-function, as the scale is almost always below the threshold of coloured
SUSY  particles (and top quark).
In
Fig.\,\ref{fig:sbdecayQCDtbmu} we can see the evolution with \tb\  and $\mu$,
which are the most interesting cases. The corrections
are large ($>10\%$) and vary slowly for large values of \tb\  
($\gsim 20$).
We remark that for 
$\mu<-120\,\GeV$ and $\tan\beta>20$ the corrections can be very large
near the phase space limit of the 
lightest sbottom decay. However, this effect
has nothing to do with the phase space exhaustion,
but rather with the presence of a dynamical
factor 
which goes to the denominator
of $\delta$ in eq.~(\ref{eq:sbtcharcorrection}). That factor is fixed by the 
structure of the interaction Lagrangian
of the sbottom decay into charginos and top; for the parameters
in Fig.\,\ref{fig:sbdecayQCDtbmu}, it turns out to vanish nearly at 
the phase space 
limit in the case of the lightest sbottom ($\tilde{b}_1$) decay.
However, this is not so either for the heaviest sbottom ($\tilde{b}_2$) or for
$\mu>120\,\GeV$ as it is patent in the same figure. The different evolution that
exhibit the corrections of the two sbottoms
has more relation with the electroweak nature of the process than with the
purely QCD  loops. For small angles 
$\theta_{\sbottom}$ and $\theta_{\stopp}$  the squarks are nearly chiral, namely
\begin{equation}
  \sbottom_1 \simeq \sbottom_R \,\,\,\,,\,\,\, \sbottom_2\simeq \sbottom_L  
  \,\,\,\,,\,\,\,
  \stopp_1 \simeq \stopp_R \,\,\,\,,\,\,\, \stopp_2\simeq \stopp_L\,\,\,, 
\label{eq:sbdecaySQchiral} 
\end{equation}
and thus their very different couplings to charginos translate
into 
very different behaviours of~(\ref{eq:sbtcharcorrection}) with \tb\  and
$\mu$. In fact, the sbottom mixing angle plays a crucial 
role,  as seen in Fig.\,\ref{fig:sbdecayQCDangle};  we
also see, however, that its value is highly constrained by 
the condition~(\ref{eq:necessary}). We should also comment on the
effect of the gaugino mass parameter $M$ and the gluino mass in
Fig.\,\ref{fig:sbdecayQCDmgM}. The gluino evolution is rather flat once the
pseudo-thresholds of $\sbottom_a\rightarrow b\,\sg$ are passed;
thus, even if the
gluino were much heavier than the squarks it would have an effect on the
sbottom decay while at the same time it would
prevent the otherwise dominant decay $\sbottom_{a}\to b\sg$.
The correction is saturated 
for the gaugino mass
parameter  $M\gsim 200\GeV$. Therefore the effects computed here can be
compared 
with the ones obtained in the higgsino approximation discussed below.
Finally, we point out the possible existence of non-decoupling effects
in the QCD part. In~\cite{DHJ} it is shown that there exist a
non-decoupling effect at large gluino masses, however this effect is
numerically small and is not the one reflected in
Fig.\,\protect{\ref{fig:sbdecayQCDmgM}}(a). The origin of the effect
is related to the breaking of SUSY, specifically to the fact that the chargino coupling 
has a renormalization group evolution which is different to that of the gauge coupling in
a non-SUSY world, the difference being sensitive to the splitting among the
various SUSY scales --
e.g.\ the scales of the squark and gluino masses.

The other parameters of the model present a rather mild effect on the
corrections for squark masses in the ballpark of several hundreds of GeV. In
summary the QCD  corrections on the decay $\sbottom_a \rightarrow t\,\cmin_i$
are large ($\simeq -20\%$ for $\sbottom_2$, $\simeq -60\%$ for $\sbottom_1$) and
negative for values of the parameter space relevant to TESLA energies,
with a higgsino-like chargino and moderate or large values of \tb.

We now turn to the discussion of the Yukawa corrections where also non-decoupling effects may come into play. They have a different origin
as compared to the pure QCD ones but they are also triggered by SUSY-breaking
and can be numerically important. We remind that, in the
computation of the Yukawa corrections, 
the higgsino approximation, eq.~(\ref{eq:defhigsi}), was used, 
and so only the lightest chargino is avaliable for the decay.
In the relevant large $\tan\beta$ segment under consideration,
namely 
\begin{equation}
20\lsim\tan\beta\lsim 40\,,
\label{eq:tansegment}
\end{equation}
the bottom quark Yukawa coupling $\lambda_b$ 
is comparable to the top quark Yukawa coupling $\lambda_t$. 
Even though the extreme interval $40<\tan\beta<60$ can be tolerated
by perturbation theory, we shall confine ourselves to the moderate range
(\ref{eq:tansegment}). This is 
necessary to preserve the condition~(\ref{eq:necessary}) for the
typical set of sparticle masses used in our analysis.

\figsbdecaytres

\figsbdecayquatre

\figsbdecaycinc

The corresponding corrections $\delta^{ai}$~(\ref{eq:sbtcharcorrection}) are shown in 
Figs.\,\ref{fig:sbdecaytres}(a) and \ref{fig:sbdecaytres}(b) as a function of the
lightest stop and sbottom masses, 
respectively. The precise value of the lightest stop mass is an important
parameter to 
determine the corrections to the lightest sbottom decay width. On the other hand the
lightest sbottom mass does not play a major role for the corrections, aside from
the presence of various thresholds.
The allowed range for the sbottom and stop mixing angles is conditioned by
the upper bound on the trilinear couplings and is obtained
from eqs.~(\ref{eq:Abt}) and~(\ref{eq:necessary}). 
In the physical
$\theta_{\sbottom}$ range, the variation
of the correction~(\ref{eq:sbtcharcorrection}) is shown in
Fig.\,\ref{fig:sbdecayquatre}(a). The large values of the corrections far away
from the allowed region~(\ref{eq:necessary}) are due to the large values of the
soft-SUSY-breaking trilinear coupling $A_b$~(\ref{eq:Abt}).
On the other hand, the permitted range for the stop mixing
angle $\theta_{\tilde{t}}$ is much larger, and we have plotted the corrections
within the allowed region in Fig.\,\ref{fig:sbdecayquatre}(b). Note that
the sign of the quantum effects for the lightest sbottom decay width changes
within the domain of variation of $\theta_{\tilde{t}}$.
Finally, we display the evolution of the SUSY-EW  
effects as a function of $\tan\beta$ (Fig.\,\ref{fig:sbdecaycinc}(a)) and of $\mu$ (Fig.\,\ref{fig:sbdecaycinc}(b))
within the region of compatibility with the constraint~(\ref{eq:necessary}).

A few words are in order to explain the origin of the leading
electroweak effects. One could expect that they come from 
the well-known large $\tan\beta$ enhancement stemming
from the chargino-stop corrections to the bottom mass  
(see e.g. Ref.~\cite{CGGJS}). 
Nonetheless this is only partially true, since in the present
case the remaining contributions 
can be sizeable enough. One can also think on the SUSY  counterpart of
the bottom mass counterterm corrections, that is, the finite contributions to
the sbottom wave function renormalization constants~\cite{GSH},
 as an additional leading contribution.  Both
of these effects are of non-decoupling nature. However the
addition of these two kind of contributions does not account for the total
behaviour in all of the parameter space.
To be more precise, in the region of the
parameter space that we have dwelled upon
the bottom mass contribution is seen to be
dominant only for the lightest sbottom decay
and for the lowest values of $\tan\beta$ in the
range~(\ref{eq:tansegment}). This is indeed the case in Fig.\,\ref{fig:sbdecayquatre}(b) where
$\tan\beta=20$ and therefore the bottom mass effect modulates the 
electroweak correction in this process and 
$\delta^{11}$ becomes essentially an odd function of the stop mixing angle. 
This fact is easily understood since, as noted above, sbottoms are nearly chiral 
-- eq.\,(\ref{eq:sbdecaySQchiral}) -- and the $\sbottom_R$ is the only one with
couples with $\lambda_b$ -- eq.\,(\ref{eq:AS}).
On the other hand, from Fig.\,\ref{fig:sbdecaycinc}(a) it is obvious that the 
(approximate) linear behaviour with $\tan\beta$ expected from bottom 
mass renormalization becomes
completely distorted by the rest of the contributions,
especially in the high $\tan\beta$ end.
In short, the final electroweak correction cannot be simply ascribed 
to a single renormalization source but to the full Yukawa-coupling combined yield.

In general the SUSY-EW   corrections to 
$\Gamma(\sbottom_a\rightarrow t\,\cmin_i)$ are smaller than 
the QCD   corrections. 
The reason why the electroweak corrections are smaller
is in part due to
the condition~(\ref{eq:necessary}) restricting our analysis within the
$\tan\beta$ interval~(\ref{eq:tansegment}). From  Figs.\,\ref{fig:sbdecayquatre} and~\ref{fig:sbdecaycinc}(a) it is
clear that outside this interval the SUSY-EW   contributions could be much higher
and with the same or opposite
sign as the QCD   effects, depending on the choice of the sign of the 
mixing angles. 
Moreover, since we have focused our analysis to sbottom masses
accessible to TESLA, again the theoretical bound~(\ref{eq:necessary})
severely restricts the maximum value of the trilinear couplings and this
prevents the electroweak corrections from being larger. 
This cannot be cured by assuming larger values of  $\mu$, because
$\mu$ directly controls the value of the (higgsino-like) chargino mass, in the final
state in the decay under study.

\section{Conclusions}

In summary, the MSSM  corrections to squark decays into charginos can be
significant and therefore must be included in any reliable 
analysis. The main corrections arise from 
the strongly interacting sector of the theory
(i.e.\ the one involving gluons and gluinos), but also
non-negligible effects may appear from the electroweak sector 
(characterized by chargino-neutralino exchange) at large (or very small)
values of $\tan\beta$. In both cases non-decoupling effects related to the 
breaking of SUSY may be involved, but it is in the electroweak part where they can
be numerically more sizeable. However, for sparticle masses of a few hundred
$GeV$ a reliable estimate of the correction requires the calculation of the
QCD and also of the complete Yukawa-coupling electroweak contribution.
The QCD  corrections  are
negative in most of the MSSM  parameter space accessible to 
TESLA\@. They are of the order
\begin{eqnarray}
  \label{eq:conclustbtcharQCD}
  \delta_{QCD} (\sbottom_1\rightarrow t\,\cmin_1) &\simeq& -60 \% \nonumber\\
  \delta_{QCD} (\sbottom_2\rightarrow t\,\cmin_1) &\simeq& -20 \% \nonumber
\end{eqnarray}
for a wide range of the parameter space (Fig.\,\ref{fig:sbdecayQCDtbmu}). In certain corners of this space, though,
they vary in a wide range of values.
EW  corrections can be of both signs. Our renormalization
prescription uses the mixing angle between squarks as an input parameter.
This prescription forces the physical region to be confined within a narrow
range when we  require compatibility with the non-existence of
colour 
breaking vacua. Within this restricted region the typical corrections
vary in the range (Figs.\,\ref{fig:sbdecayquatre}, \ref{fig:sbdecaycinc})
\begin{eqnarray}
  \label{eq:conclusstbcharEW}
  \delta_{EW} (\sbottom_1\rightarrow t\,\cmin_1) &\simeq& +25 \% \mbox{ to } -15 \% \nonumber\\
  \delta_{EW} (\sbottom_2\rightarrow t\,\cmin_1) &\simeq& +5 \% \mbox{ to } -5 \% \,\,,\nonumber
\end{eqnarray}
However we must recall  that these limits are
not rigorous. In the edge of such regions  we find the largest EW
contributions. 
We stress that for these decays
it is not possible to narrow down the bulk of the electroweak corrections to just some
simple-structured leading terms.

The present study has an impact on the determination of squark parameters
at TESLA\@. The squark masses used in it would be available already for TESLA running
at a center of mass energy of $800\GeV$. The large corrections  found from both the
QCD  and the EW (Yukawa) sector, make this calculation necessary,
not only for prospects of precision measurements in the
sbottom-chargino-neutralino sectors, but also for a reliable first determination
of their parameters.

\section*{Acknowledgments}
This work has been partially supported by the Deutsche Forschungsgemeinschaft
and by CICYT under project No. AEN99-0766.

\end{document}